\documentclass[pra,showpacs,showkeys,amsfonts,amsmath,twocolumn]{revtex4}
\usepackage{bm}
\usepackage{graphicx}
\RequirePackage{mathptm}

\numberwithin{equation}{section}
\newcommand{\si}[1]{\sigma_{#1}}

\newcommand{\sa}[2]{\sigma_{#1}^{#2}}

\newcommand{\ro}{\varrho}

\newcommand{\al}[1]{\alpha_{#1}}
\newcommand{\fal}[1]{\widetilde{\alpha}_{#1}}

\newcommand{\ga}[1]{\gamma_{#1}}

\newcommand{\te}{\theta}

\newcommand{\I}{\openone}
\newcommand{\conj}[1]{\overline{#1}}
\newcommand{\fala}[1]{\widetilde{#1}}
\newcommand{\ket}[1]{|{#1}\rangle}
\newcommand{\bra}[1]{\langle {#1} |}

\newcommand{\R}{\mathbb R}

\newcommand{\PP}{\mathbb P}

\newcommand{\tr}{\mathrm{tr}\,}
\newcommand{\ptr}[1]{\mathrm{tr}_{#1}}
\newcommand{\tl}[1]{\boldsymbol #1}
\newcommand{\mr}[1]{\mathrm{#1}}

\newcommand{\DS}{\displaystyle}

\begin{document}
\title{Spontaneous emission and quantum discord: comparison of Hilbert-Schmidt and trace distance discord}
\author{  Lech Jak{\'o}bczyk
\footnote{ ljak@ift.uni.wroc.pl} }
 \affiliation{Institute of Theoretical Physics\\ University of
Wroc{\l}aw\\
Plac Maxa Borna 9, 50-204 Wroc{\l}aw, Poland}
\begin{abstract}
Hilbert - Schmidt  and trace norm geometric quantum discord are
compared with regard to their behavior during local time evolution.
We consider the system of independent two - level atoms with  time
evolution  given by the dissipative process of spontaneous emission.
It is explicitly shown that the Hilbert - Schmidt norm discord has
nonphysical properties with respect to such local evolution and
cannot serve as a reasonable measure of quantum correlations and the
better choice is to use trace norm discord as such a measure.
\end{abstract}
\maketitle
\section{Introduction}
Characterizing the nature of correlations in composite quantum
systems is one of the fundamental problems in quantum theory. When
the system is prepared in a pure state, only entanglement is
responsible for the presence of quantum correlations. On the other
hand, once mixed states are taken into account, the problem becomes
much more involved. Some features of separable mixed states are
incompatible with a classical description of correlations. The most
important among them is that a measurement on a part of composite
system in some non-entangled states can induce disturbance on the
state of complementary subsystem. Such "non-classical" behavior can
be quantified by quantum discord - the most promising measure of
bipartite quantum correlations beyond quantum entanglement \cite{Z}.
For pure states discord coincides with entanglement, but in the case
of mixed states discord and entanglement differ significantly. For
example it was shown that almost all quantum states have
non-vanishing discord \cite{F} and  even local operations on the
measured part can increase or create quantum discord \cite{Str,
Hu1}.
\par
In this paper we quantify non-classical correlations which may
differ from entanglement by using geometric quantum discord. This
quantity is defined in terms of minimal distance of the given state
from the set of classically - correlated states, so the proper
choice of such a distance is crucial. The measure proposed in
\cite{DVB} uses a Hilbert - Schmidt norm to define a distance in the
set of states. This choice has a technical advantage: the
minimization process can be realized analytically for arbitrary
two-qubit states. Despite of this feature, this measure has some
unwanted properties. The most important problem is that it may
increase under local operations performed on the unmeasured
subsystem \cite{Piani, Tuf}. Fortunately, by using other norm in the
set of states, this defect can be repaired: the best choice is to
use Schatten 1-norm (or trace norm) to define quantum discord
\cite{Paula}. On the other hand, such defined measure is more
difficult to compute. The closed formula for it is known only in the
case of Bell - diagonal states or X - shaped two - qubit states
\cite{Paula, Cic}.
\par
The main scope of this paper is to reconsider the properties of those
two measures of quantum discord in a concrete physical system where the 
quantum channel is given by the time evolution. As a compound system 
we take two independent two - level atoms not completely isolated from the environment.
In this case the time evolution  is given by a dissipative
process of spontaneous emission. One - sided spontaneous emission in
which only one atom emits photons and the other is isolated from the
environment, gives the physical realization of local quantum
channel. Although it was already established \cite{Piani, Tuf}, in this framework 
we can  explicitly show that Hilbert - Schmidt norm discord
has  nonphysical properties with respect to the local evolution and the 
better choice is to use trace norm. In particular we discuss the local creation of
discord when the system is prepared in classical initial state
\cite{Ci,Ci1,Ge,Ca}.  In Ref.\cite{GJ} we have studied time evolution of Hilbert
- Schmidt quantum discord $D_{2}$, now we compare it with the
behavior of trace norm quantum discord $D_{1}$. The results of our
analysis show that when only the  local creation of quantum discord
in the classical initial state is considered, $D_{2}$ and $D_{1}$
provide the same information about the evolution of quantum
correlations. This is no longer true when the initial states have
non - zero discord. Local evolution can increase quantum discord and
this phenomenon can be observed  by using $D_{1}$ or $D_{2}$. On the
other hand, there are initial states with decreasing quantum
correlations quantified by $D_{1}$ whereas $D_{2}$ is increasing.
The most spectacular manifestation of nonphysical properties of
Hilbert - Schmidt norm discord is its behavior during the local
evolution of the unmeasured subsystem. $D_{2}$ not only increases
for a large class of initial discordant states (at the same time
$D_{1}$ obviously decreases) but also it can increase even when the
local evolution of the measured subsystem leads to decreasing
$D_{2}$. This shows again that in contrast to trace norm discord,  
Hilbert - Schmidt norm discord cannot serve as a reasonable measure of quantum
correlations.
\section{Geometric measures of quantum discord}
We start with the introduction of the standard notion
of geometric quantum discord \cite{DVB}. When a $d\otimes d$
bipartite system $AB$ is prepared in a state $\ro$ and we perform
local measurement on the subsystem $A$, almost all  states $\ro$
will be disturbed due to such measurement. The (one-sided) geometric
discord $D_{2}(\ro)$ can be defined as the minimal disturbance,
measured by the squared Hilbert-Schmidt distance, induced by any
projective measurement $\PP^{A}$ on subsystem $A$ i.e.
\begin{equation}
D_{2}(\ro)=\frac{d}{d-1}\;\min\limits_{\PP^{A}}\,||\ro-\PP^{A}(\ro)||_{2}^{2},
\label{DG}
\end{equation}
where
\begin{equation}
||a||_{2}=\sqrt{\tr a\,a^{\ast}}.
\end{equation}
Here we adopt normalized version of the geometric discord,
introduced in Ref. \cite{Adesso}. In the case of two qubits, there is an
explicit expression for $D_{2}$ \cite{DVB}:
\begin{equation}
D_{2}(\ro)=\frac{1}{2}\,\left(||\tl{x}||^{2}+||T||_{2}^{2}-k_{\mathrm{max}}\right),
\end{equation}
where the components of the vector $\tl{x}\in \R^{3}$ are given by
\begin{equation}
 x_{k}=\tr\, (\ro\,\si{k}\otimes \I),
\end{equation}
the matrix $T$ has elements
\begin{equation}
T_{jk}=\tr\,(\ro\,\si{j}\otimes \si{k})
\end{equation}
and $k_{\mathrm{max}}$ is the largest eigenvalue of the matrix
$\tl{x}\,\tl{x}^{T}+T\,T^{T}$. Despite of being easy to compute, the
measure $D_{2}$ fails as a quantifier of quantum correlations, since
it may increase under local operations on the unmeasured subsystem
\cite{Piani}. In the present paper we explicitly show that one-sided
spontaneous emission of the unmeasured atom can create additional
discord quantified by $D_{2}$ in the system of two independent
atoms. Such defect of $D_{2}$ originates in the properties of
Hilbert -Schmidt norm, which manifests also in the case of
entanglement \cite{Ozawa}.
\par
To repair this defect, one considers  other norms in the set of quantum states. The
best choice is to use the trace norm (or Schatten 1-norm)  and define \cite{Paula}
\begin{equation}
D_{1}(\ro)=\min\limits_{\PP^{A}}\,||\ro-\PP^{A}(\ro)||_{1},
\end{equation}
where
\begin{equation}
||a||_{1}=\tr\, |a|.
\end{equation}
$D_{1}$ has desired properties with respect to the local operations on unmeasured
subsystem, but its computation is much more involved. Analytic expression for $D_{1}$
is known only for limited classes of two - qubits states, including Bell - diagonal \cite{Paula}
and $X$ - shaped mixed states \cite{Cic}. In the present paper, we consider $X$ - shaped two - qubit
states
\begin{equation}
\ro=\begin{pmatrix}\ro_{11}&0&0&\ro_{14}\\
0&\ro_{22}&\ro_{23}&0\\
0&\ro_{32}&\ro_{33}&0\\
\ro_{41}&0&0&\ro_{44}
\end{pmatrix},\label{X}
\end{equation}
where all matrix elements are real and non - negative. The quantity $D_{1}$ for such states can be computed
as follows. Let  $x=2(\ro_{11}+\ro_{22}) -1$ and
\begin{equation}
\al{1}=2(\ro_{23}+\ro_{14}),\quad \al{2}=2(\ro_{23}-\ro_{14}),\quad \al{3}=1-2(\ro_{22}+\ro_{33}).
\end{equation}
Then \cite{Cic}
\begin{equation}
D_{1}(\ro)=\sqrt{\frac{\DS a\,\al{1}^{2}-b\,\al{2}^{2}}{\DS a-b +\al{1}^{2}-\al{2}^{2}}},\label{disc1}
\end{equation}
where
\begin{equation}
a=\max\,(\al{3}^{2},\, \al{2}^{2}+x^{2}),\quad b=\min\,(\al{3}^{2},\, \al{1}^{2}).
\end{equation}
Notice that we use normalized version of $D_{1}$ and the formula
(\ref{disc1}) is not valid in the case when $x=0$ and
\begin{equation}
|\al{1}|=|\al{2}|=|\al{3}|.
\end{equation}
In such a case, one can use general prescription how to compute
$D_{1}$, also given in Ref. \cite{Cic} (eq. (65)).
\par
In the case of pure states, $D_{1}$ as well as $D_{2}$ give the same
information about quantum correlations as entanglement measured by
negativity
\begin{equation}
N(\ro)=||\ro^{\mr{PT}}||_{1}-1,
\end{equation}
where $\ro^{\mr{PT}}$ denotes partial transposition of $\ro$. In the case of mixed
states, entanglement and discord significantly differ. For example
 for two - qubit Bell - diagonal states one finds that
\cite{Paula}
\begin{equation}
D_{1}\geq \sqrt{D_{2}}\geq N.\label{d1d2}
\end{equation}
The inequality $\sqrt{D_{2}}\geq N$ was proved to be valid for all
two - qubit mixed states \cite{Adesso}, and it is conjectured that
(\ref{d1d2}) is also valid for all  two - qubit states.
\par
To show that inequalities in (\ref{d1d2}) can be sharp, consider the
following family of states \cite{Mizra}
\begin{equation}
\ro_{\te}=\begin{pmatrix}\frac{1}{2}\cos^{2}\te&0&0&\frac{1}{4}\sin 2\te\\[2mm]
0&0&0&0\\[2mm]
0&0&\frac{1}{2}&0\\[2mm]
\frac{1}{4}\sin 2\te&0&0&\frac{1}{2}\sin^{2}\te
\end{pmatrix},\label{rote}
\end{equation}
where $\te \in [0,\pi/2]$. By direct computation, one can check that
\begin{equation}
N(\ro_{\te})=\frac{\DS \sqrt{6-2\cos 4\te}-2}{\DS 4},
\end{equation}
whereas
\begin{equation}
D_{2}(\ro_{\te})=\min\,\left(\frac{1}{2}\sin^{2}\te,\, \frac{1}{4}\sin^{2} 2\te\right)
\end{equation}
and
\begin{equation}
D_{1}(\ro_{\te})=\frac{1}{2}\sin 2\te.
\end{equation}
Notice that
\begin{equation}
D_{1}(\ro_{\te})> \sqrt{D_{2}(\ro_{\te})}> N(\ro_{\te}
\end{equation}
if $\te\in (0,\pi/4)$ and
\begin{equation}
D_{1}(\ro_{\te})=\sqrt{D_{2}(\ro_{\te})}>N(\ro_{\te})
\end{equation}
for $\te \in [\pi/4,\pi/2]$ (see FIG.1).
\begin{figure}[h]
\centering {\includegraphics[height=54mm]{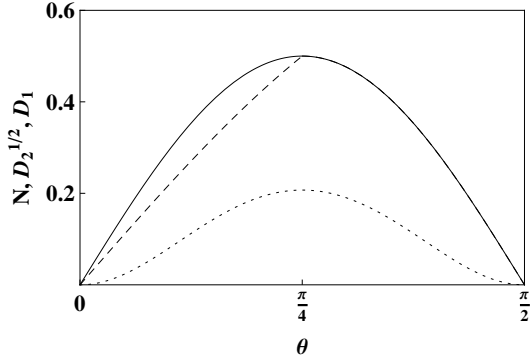}}\caption{$N$
(doted line), $\sqrt{D_{2}}$ (dashed line) and $D_{1}$ (solid line)
as a functions of the parameter $\theta$ for the states
(\ref{rote}).}
\end{figure}
\section{Local dynamics  induced by spontaneous emission and geometric discord}
\subsection{One - sided spontaneous emission}
Consider a system of two independent two - level atoms (atom $A$ and
atom $B$) interacting with environment at zero temperature. In this study we
take into account only the dissipative process of spontaneous
emission, so  the dynamics of the system is given by the master equation
\cite{Agarwal}
\begin{equation}
\frac{d\ro}{dt}=L_{AB}\ro,\quad L_{AB}=L_{A}+L_{B},\label{me}
\end{equation}
where for $k=A,\, B$
\begin{equation}
L_{k}=\frac{\ga{0}}{2}\,\left(2\,\sa{-}{k}\ro\sa{+}{k}-\sa{+}{k}\sa{-}{k}\ro-\ro\sa{+}{k}\sa{-}{k}\right).
\end{equation}
In the above equation $\sa{\pm}{A}=\si{\pm}\otimes \I,\,
\sa{\pm}{B}=\I\otimes \si{\pm}$ and $\ga{0}$ is the single atom
spontaneous emission rate. Local evolution of the atom $A$ is given by "one-sided" spontaneous
emission generated only by the generator $L_{A}$ i.e.
\begin{equation}
\ro_{t,A}=T_{t}^{A}\ro,\quad T_{t}^{A}=e^{tL_{A}}.\label{leA}
\end{equation}
In this case the atom $A$ spontaneously emits photons, whereas
the atom $B$ is isolated from the environment. Similarly we can consider
one-sided spontaneous emission of the atom $B$ i.e. the evolution
\begin{equation}
\ro_{t,B}=T_{t}^{B}\ro,\quad T_{t}^{B}=e^{tL_{B}}.\label{leB}
\end{equation}
In what follows we consider the $X$ - shaped initial states (\ref{X}),
where the matrix elements of $\ro$ are given
 with respect to the basis
$\ket{e}_{A}\otimes
\ket{e}_{B},\ket{e}_{A}\otimes \ket{g}_{B}, \ket{g}_{A}\otimes
\ket{e}_{B}, \ket{g}_{A}\otimes \ket{g}_{B}$
 and $\ket{g}_{k},\ket{e}_{k},\, k=A,B$ are the ground states and excited
states of atoms $A$ and $B$.
For such initial state, the state $\ro_{t,A}$ has the following matrix elements
\begin{equation}
\begin{split}
& (\ro_{t,A})_{11}=e^{-\ga{0}t}\ro_{11},\\
& (\ro_{t,A})_{22}=e^{-\ga{0}t}\ro_{22},\\
& (\ro_{t,A})_{33}=(1-e^{-\ga{0}t})\ro_{11}+\ro_{33},\\
& (\ro_{t,A})_{44}=(1-e^{-\ga{0}t})\ro_{22}+\ro_{33},\\
& (\ro_{t,A})_{14}=e^{-\ga{0}t/2}\ro_{14},\\
& (\ro_{t,A})_{23}=e^{-\ga{0}t/2}\ro_{23}.
\end{split}
\end{equation}
Similarly
\begin{equation}
\begin{split}
& (\ro_{t,B})_{11}=e^{-\ga{0}t}\ro_{11},\\
& (\ro_{t,A})_{22}=(1-e^{-\ga{0}t})\ro_{11}+\ro_{22},\\
& (\ro_{t,A})_{33}=e^{-\ga{0}t}\ro_{33},\\
& (\ro_{t,A})_{44}=(1-e^{-\ga{0}t})\ro_{33}+\ro_{44},\\
& (\ro_{t,A})_{14}=e^{-\ga{0}t/2}\ro_{14},\\
& (\ro_{t,A})_{23}=e^{-\ga{0}t/2}\ro_{23}.
\end{split}
\end{equation}
Notice that in contrast to the usual process of spontaneous emission,
for the the one-sided emissions, there are non-trivial asymptotic states:
one can check that for any initial state $\ro$ when $t\to \infty$
\begin{equation}
\ro_{t,A}\to P_{g}^{A}\otimes \ptr{A}\,\ro
\end{equation}
and
\begin{equation}
\ro_{t,B}\to \ptr{B}\,\ro \otimes P_{g}^{B}.
\end{equation}
where $P_{g}^{A},\, P_{g}^{B}$ are projections on the ground states
of the atom $A$ and  $B$ respectively.
\subsection{Time evolution of $D_{1}$ and $D_{2}$}
Now we study quantum correlations in the states $\ro_{t,A}$ and $\ro_{t,B}$ defined above.
We start with trace distance geometric discord. In the state $\ro_{t,A}$ we have
\begin{equation}
D_{1}(\ro_{t,A})=\sqrt{\frac{\DS a(t)\,\al{1}(t)^{2}-b(t)\,\al{2}(t)^{2}}
{\DS a(t)-b(t) +\al{1}(t)^{2}-\al{2}(t)^{2}}},\label{disc1A}
\end{equation}
where
\begin{equation}
\begin{split}
&\al{1}(t)=2(\ro_{14}+\ro_{23})\,e^{-\ga{0}t/2},\\
&\al{2}(t)=2(\ro_{23}-\ro_{14})\,e^{-\ga{0}t/2},\\
&\al{3}(t)=2(\ro_{11}-\ro_{22})\,e^{-\ga{0}t} -2(\ro_{11}+\ro_{33})+1,\\
&x(t)=\,2(\ro_{11}+\ro_{22})\,e^{-\ga{0}t} -1
\end{split}
\end{equation}
and
\begin{equation}
\begin{split}
&a(t)=\max\,(\al{3}(t)^{2},\, \al{2}(t)^{2}+x(t)^{2}),\\
& b(t)=\min\,(\al{3}(t)^{2},\, \al{1}(t)^{2}).
\end{split}
\end{equation}
Similarly
\begin{equation}
D_{1}(\ro_{t,B})=\sqrt{\frac{\DS \fala{a}(t)\,\fal{1}(t)^{2}-\fala{b}(t)\,\fal{2}(t)^{2}}
{\DS \fala{a}(t)-\fala{b}(t) +\fal{1}(t)^{2}-\fal{2}(t)^{2}}},\label{disc1B}
\end{equation}
where
\begin{equation}
\begin{split}
&\fal{1}(t)=\al{1}(t),\quad \fal{2}(t)=\al{2}(t),\quad \fala{x}(t)=x,\\
&\fal{3}(t)=2(\ro_{11}-\ro_{33})\,e^{-\ga{0}t} -2(\ro_{11}+\ro_{22})+1
\end{split}
\end{equation}
and
\begin{equation}
\begin{split}
&\fala{a}(t)=\max\,(\fal{3}(t)^{2},\, \fal{2}(t)^{2}+\fala{x}(t)^{2}),\\
&\fala{b}(t)=\min\,(\fal{3}(t)^{2},\, \fal{1}(t)^{2}).
\end{split}
\end{equation}
Concerning $D_{2}$, one finds
\begin{equation}
D_{2}(\ro_{t,A})=\min\,\left(f_{1}(t),\, f_{2}(t),\, f_{3}(t)\right),\label{D2X}
\end{equation}
where
\begin{equation}
\begin{split}
&f_{1}(t)=4(\ro_{14}^{2}+\ro_{23}^{2})\,e^{-\ga{0}t},\\[2mm]
&f_{2}(t)=4(\ro_{11}^{2}+\ro_{22}^{2})\,e^{-2\ga{0}t}+2\big[(\ro_{14}-\ro_{23})^{2}\\
&\hspace*{10mm}-2\ro_{11}(\ro_{11}+\ro_{33})-2\ro_{22}(\ro_{22}+\ro_{44})\big]\,e^{-\ga{0}t}\\
&\hspace*{10mm}+(\ro_{11}+\ro_{33})^{2}+(\ro_{22}+\ro_{44})^{2},\\[2mm]
&f_{3}(t)=4(\ro_{11}^{2}+\ro_{22}^{2})\,e^{-2\ga{0}t}+2\big[(\ro_{14}+\ro_{23})^{2}\\
&\hspace*{10mm}-2\ro_{11}(\ro_{11}+\ro_{33})-2\ro_{22}(\ro_{22}+\ro_{44})\big]\,e^{-\ga{0}t}\\
&\hspace*{10mm}+(\ro_{11}+\ro_{33})^{2}+(\ro_{22}+\ro_{44})^{2}.
\end{split}
\end{equation}
Similarly
\begin{equation}
D_{2}(\ro_{t,B})=\min\,\left(\fala{f}_{1}(t),\, \fala{f}_{2}(t),\, \fala{f}_{3}(t)\right)\label{d2B}
\end{equation}
where
\begin{equation}
\begin{split}
&\fala{f}_{1}(t)=f_{1}(t),\\[2mm]
&\fala{f}_{2}(t)=2(\ro_{11}-\ro_{33})^{2}\,e^{-2\ga{0}t}+2\big[ (\ro_{14}-\ro_{23})^{2}\\
&\hspace*{10mm}-(\ro_{11}-\ro_{33})(\ro_{11}+\ro_{22}-\ro_{33}-\ro_{44})\big]\,e^{-\ga{0}t}\\
&\hspace*{10mm}+(\ro_{11}+\ro_{22})^{2}+(\ro_{33}+\ro_{44})^{2}\\
&\hspace*{10mm}-2(\ro_{11}+\ro_{22})(\ro_{33}+\ro_{44}),\\[2mm]
&\fala{f}_{3}(t)=2(\ro_{11}-\ro_{33})^{2}\,e^{-2\ga{0}t}+2\big[ (\ro_{14}+\ro_{23})^{2}\\
&\hspace*{10mm}-(\ro_{11}-\ro_{33})(\ro_{11}+\ro_{22}-\ro_{33}-\ro_{44})\big]\,e^{-\ga{0}t}\\
&\hspace*{10mm}+(\ro_{11}+\ro_{22})^{2}+(\ro_{33}+\ro_{44})^{2}\\
&\hspace*{10mm}-2(\ro_{11}+\ro_{22})(\ro_{33}+\ro_{44}).
\end{split}
\end{equation}
\subsection{Classically correlated initial states}
 We choose as initial states the following $X$ -shaped states
\begin{equation}
\ro_{c}=\begin{pmatrix} w&0&0&s\\
0&\frac{1}{2}-w&s&0\\
0&s&w&0\\
s&0&0&\frac{1}{2}-w\end{pmatrix},\label{roc}
\end{equation}
where
\begin{equation}
0<w<\frac{1}{2}, \quad 0<s\leq s_{\mr{max}}
\end{equation}
and
\begin{equation}
s_{\mr{max}}=\sqrt{\frac{1}{2}w-w^{2}}.
\end{equation}
One can check that
\begin{equation}
D_{1}(\ro_{c})=D_{2}(\ro_{c})=0,
\end{equation}
so $\ro_{c}$ are only classically correlated. Notice that for such initial states
\begin{equation}
\begin{split}
&\al{1}(t)=4s\,e^{-\ga{0}t/2},\quad \al{2}(t)=0\\
&\al{3}(t)=(1-4w)(1-e^{-\ga{0}t}),\quad x(t)=e^{-\ga{0}t}-1,
\end{split}
\end{equation}
so
\begin{equation}
a(t)=\max\,\left(\al{3}(t)^{2},x(t)^{2}\right)=\al{3}(t)^{2}
\end{equation}
and
\begin{equation}
D_{1}(\ro_{t,A})=\frac{\DS 4s\,(1-e^{-\ga{0}t})}{G(t)},\label{genD1}
\end{equation}
where
\begin{equation}
G(t)=\sqrt{16s^{2}+g(t)-\min (16s^{2}, g(t)(1-4w)^{2})}
\end{equation}
and
\begin{equation}
g(t)=2\,(\cosh\ga{0}t -1).
\end{equation}
One can check that (\ref{genD1}) as a function of $t$ grows from zero to some maximal value
and then asymptotically vanishes. So for any initial state (\ref{roc}) there is a local
generation of transient quantum correlations measured by $D_{1}$. The most efficient production
of discord is when $w=1/4$ and in that case, the maximum is achieved for $s=1/4$ i.e. for initial
state of the form
\begin{equation}
\ro_{0}=\frac{1}{2}\,\ket{+}\bra{+}\otimes
\ket{+}\bra{+}\,+\,\frac{1}{2}\,\ket{-}\bra{-}\otimes
\ket{-}\bra{-}, \label{roo}
\end{equation}
where
\begin{equation}
\ket{\pm}=\frac{1}{\sqrt{2}}\, \left(\ket{e}\pm \ket{g}\right).
\end{equation}
Due to the properties of trace distance, $D_{1}$ is non- increasing
under general local operations on subsystem $B$, so it is equal to
zero for all $t$. In our  model of local evolutions we can check it
explicitly:  for initial states (\ref{roc})
 $\fal{1}(t)=4s\,e^{-\ga{0}t/2} $ but
 $\fal{2}(t)=\fal{3}(t)=\fala{x}(t)=0$, so $D_{1}(\ro_{t,B})=0$.
\par
Now we consider the same problem, but using Hilbert - Schmidt distance discord $D_{2}$. In the case of
initial state $\ro_{c}$ we have
\begin{equation}
\begin{split}
&f_{1}(t)=8s^{2}\,e^{-\ga{0}t},\\
&f_{2}(t)=(1-4w+8w^{2})(1-e^{-\ga{0}t})^{2},\\
&f_{3}(t)=(1-4w+8w^{2})(1-e^{-\ga{0}t})^{2}+8s^{2}\,e^{-\ga{0}t}.
\end{split}
\end{equation}
Notice that $f_{3}(t)>f_{2}(t)$ and $f_{1}(t)$ is decreasing from the value $8s^{2}$  to zero,
whereas $f_{2}(t)$ is increasing from zero to the value $1-4w+8w^{2}$, as $t$ goes to infinity.
So there is  the time $t_{\mr{max}}$ at which those functions are equal and $D_{2}$ defined by
the formula (\ref{D2X}) grows from zero to some maximal value and then asymptotically vanishes.
Similarly as in the case of $D_{1}$, maximal production of discord is for the initial state
(\ref{roo}). Moreover, since $\fala{f}_{2}(t)=0$, $D_{2}(\ro_{t,B})=0$. Thus we have shown that
 as far as the process of local generation of quantum correlations out of classical correlations
is concerned, $D_{1}$ and $D_{2}$ give the similar information (see FIG.2).
\begin{figure}[h]
\centering {\includegraphics[height=54mm]{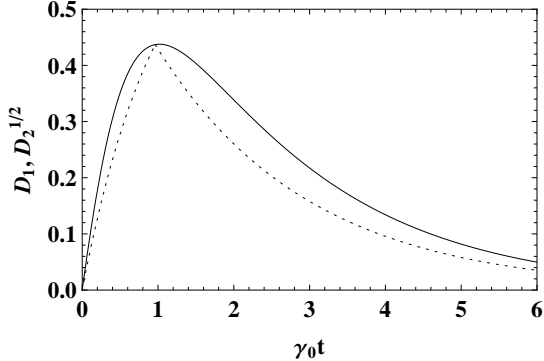}}\caption{Time
evolution of $D_{1}$ (solid line) and $\sqrt{D_{2}}$ (dotted line)
for the initial state $\ro_{0}$, under one - sided emission  of the
atom $A$.}
\end{figure}
\subsection{Some initial states with non - zero discord}
Local evolution can also increase the initial non - zero discord. To show this in our model,
consider the  states of the form
\begin{equation}
\ro_{d}=\begin{pmatrix} w&0&0&s\\0&w&s&0\\0&s&\frac{1}{2}-w&0\\
s&0&0&\frac{1}{2}-w\end{pmatrix},\label{rod}
\end{equation}
where as in the case of $\ro_{c}$
\begin{equation}
0<w<\frac{1}{2}\quad\text{and}\quad 0<s\leq s_{\mr{max}}.
\end{equation}
In contrast to the states (\ref{roc}), the states $\ro_{d}$  have non - zero discord, for all $w\in (0,1/2)$
(except of $w=1/4$) and admissible $s$. One can check that
\begin{equation}
D_{1}(\ro_{d})=\frac{\DS 4s\,|1-4w|}{\DS \sqrt{16s^{2}+(1-4w)^{2}}}
\end{equation}
and
\begin{equation}
D_{2}(\ro_{d})=\min\,\left( 8s^{2},\, 2\left(2w-\frac{1}{2}\right)^{2}\right).
\end{equation}
\par
We start the analysis of time evolution of quantum correlations by considering first
Hilbert - Schmidt discord $D_{2}$. For the local spontaneous emission of the atom $A$
and initial state $\ro_{d}$ we have
\begin{equation}
\begin{split}
&f_{1}(t)=8s^{2}\,e^{-\ga{0}t},\\
&f_{2}(t)=\frac{1}{2}-4w\,e^{-\ga{0}t} +8w^{2}\,e^{-2\ga{0}t},\\
&f_{3}(t)=\frac{1}{2}-4w\,e^{-\ga{0}t} +8w^{2}\,e^{-2\ga{0}t}+8s^{2}\,e^{-\ga{0}t}.
\end{split}
\end{equation}
Since $f_{3}(t)> f_{2}(t)$, only the relations between $f_{1}(t)$
and $f_{2}(t)$ are crucial for the behavior of $D_{2}$. Notice that
$f_{1}(t)$ decreases form the value $8s^{2}$ and goes to zero. On the
other hand, the function $f_{2}(t)$ may be increasing or decreasing,
depending on the value of the parameter $w$. It can be shown that if
$ 0<w\leq 1/4$, $f_{2}(t)$ increases, whereas if $1/4<w<1/2$,
$f_{2}(t)$ initially decreases and then start to increase. The
production of an additional discord can happen only when $f_{2}$ is
increasing function of $t$ and  when $f_{1}(0)>f_{2}(0)$ i.e. when
$0<w<1/4$ and
\begin{equation}
8s^{2}>\frac{1}{2}-4w+8w^{2}. \label{cond}
\end{equation}
 The condition (\ref{cond}) gives some restrictions on the possible
values of the parameter $w$ in order to obtain production of
discord. If we put $s_{\mr{max}}$ into the inequality (\ref{cond}),
 we obtain the critical value of $w$ given by
\begin{equation}
w_{c}=\frac{1}{8}(2-\sqrt{2}).
\end{equation}
One can show that if $0<w\leq w_{c}$,  each pair $(w,s)$ where
$0<s\leq s_{\mr{max}}$ defines the initial state with only
decreasing $D_{2}$. On the other hand, if $w_{c}<w<1/4$, a pair
$(w,s_{\mr{max}})$, gives the initial state
with increasing discord. The values of $w$ greater then $1/4$ always
give decreasing $D_{2}$.
\par
Now we consider trace norm discord $D_{1}$. For the initial state
$\ro_{d}$ we obtain
\begin{equation}
\begin{split}
&\al{1}(t)=4s\,e^{-\ga{0}t/2},\quad \al{2}(t)=\al{3}(t)=0,\\
&x(t)=4w\,e^{-\ga{0}t}-1,
\end{split}
\end{equation}
so
\begin{equation}
D_{1}(\ro_{t,A})=\frac{\DS 4s\,|1-4w\,e^{-\ga{0}t}|}{\DS
\sqrt{16s^{2}+e^{\ga{0}t}\,(1-4w\,e^{-\ga{0}t})^{2}}}.\label{d1d}
\end{equation}
Since the numerator of the right hand side of (\ref{d1d}) increases
only when $0<w\leq 1/4$ and denominator always increases, for such
values of $w$, $D_{1}$ may increase for some period of time.
Detailed analysis of the formula (\ref{d1d}) shows that similarly as
in the case of $D_{2}$, there is the critical value $\conj{w}_{c}$
such that for any pair $(w,s),\, 0<w\leq \conj{w}_{c}$ and
admissible $s$, the corresponding initial state gives decreasing
discord $D_{1}$, whereas any pair $(w,s_{\mr{max}}),\,
\conj{w}_{c}<w<1/4$, defines initial state with growing  $D_{1}$.
The crucial for this analysis is the fact that $\conj{w}_{c}$ is
slightly larger then $w_{c}$ ($\conj{w}_{c}\approx 0.0777$ and
$w_{c}\approx 0.0732$), so there are the initial states
corresponding to $w\in (w_{c},\, \conj{w}_{c})$ such that two
measures of geometric discord behave very differently: $D_{2}$ grows
for some period of time, whereas $D_{1}$ decreases for all $t$ (see
FIG.3).
\begin{figure}[t]
\centering {\includegraphics[height=54mm]{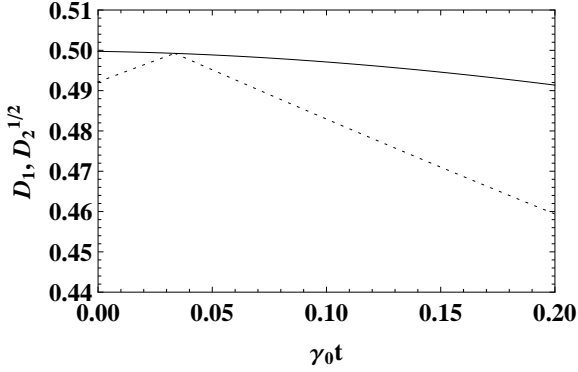}}\caption{Time evolution of $D_{1}$ (solid line) and $\sqrt{D_{2}}$ (dotted line)
for the initial state $\ro_{d}$ with $w=0.076,\, s=0.179$.}
\end{figure}
For the initial states with values of $w$ between $\conj{w}_{c}$ and $1/4$, two measures of discord behave similarly:
for a finite period of time, $D_{1}$ and $D_{2}$ grow to some maximal value and then start to decrease (FIG.4).
\begin{figure}[b]
\centering {\includegraphics[height=54mm]{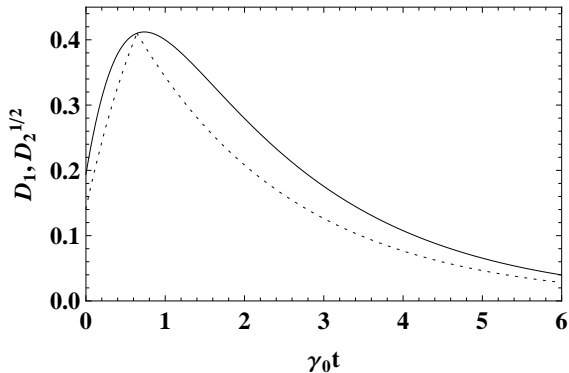}}\caption{Time evolution of $D_{1}$ (solid line) and $\sqrt{D_{2}}$ (dotted line)
for the initial state $\ro_{d}$ with $w=0.2,\, s=0.2$.}
\end{figure}
Notice also that for the initial states with $w>1/4$ the function (\ref{d1d}) decreases to zero at finite
time $t_{0}=(\ln 4w)/\ga{0}$ and then starts to grow to some maximal value. The similar behavior can be observed also
in the case of $D_{2}$ (FIG.5).
\begin{figure}[t]
\centering {\includegraphics[height=54mm]{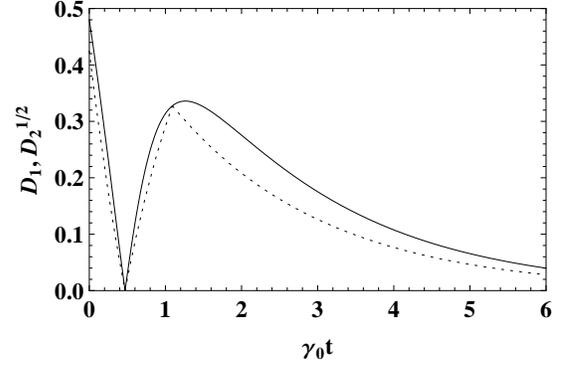}}\caption{Time evolution of $D_{1}$ (solid line) and $\sqrt{D_{2}}$ (dotted line)
for the initial state $\ro_{d}$ with $w=0.4,\, s=0.2$.}
\end{figure}
\par
Now we consider time evolution given by local spontaneous emission
of atom $B$. Since $D_{1}$ is non - increasing under local
operations on subsystem $B$, we only consider the properties of
$D_{2}(\ro_{t,B})$. It is given by formula (\ref{d2B}), where
\begin{equation}
\begin{split}
&\fala{f}_{1}(t)=8s^{2}\,e^{-\ga{0}t},\\
&\fala{f}_{2}(t)=c-c\,e^{-\ga{0}t}+\frac{1}{2}c\, e^{-2\ga{0}t},\\
&\fala{f}_{3}(t)=c-c\,e^{-\ga{0}t}+\frac{1}{2}c\, e^{-2\ga{0}t}+8s^{2}\,e^{-\ga{0}t}
\end{split}
\end{equation}
and
\begin{equation}
c=1-8w+16w^{2}.
\end{equation}
\begin{figure}[b]
\centering {\includegraphics[height=54mm]{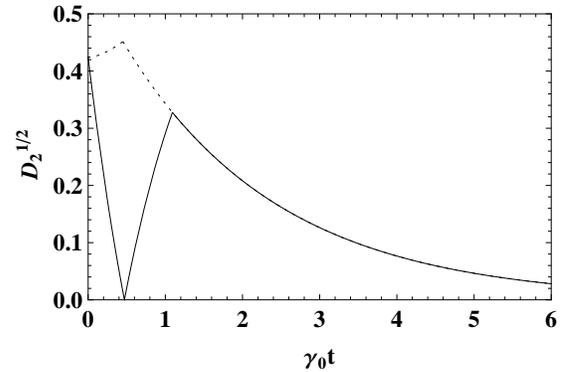}}\caption{Time
evolution of $\sqrt{D_{2}}$ given by local emission of atom $A$
(solid line) versus local emission of atom $B$  (dotted line), for
the initial state $\ro_{d}$ with $w=0.4,\, s=0.2$.}
\end{figure}
One can check that in this case the function $\fala{f}_{2}(t)$ is
increasing for all values of $w$ (except of $w=1/4$) and the
condition that $\fala{f}_{1}(0)>\fala{f}_{2}(0)$ is the same as in
the case of evolution of atom $A$. Thus for all pairs $(w,\,
s_{\mr{max}})$ where $w\in (w_{c},\, 1/2],\; w\neq 1/4$, the
corresponding initial state gives increasing $D_{2}$ under the local
evolution of unmeasured subsystem $B$. This happens even in the case
when local evolution of measured atom $A$ leads to decreasing
$D_{2}$. It explicitly shows  nonphysical properties of Hilbert -
Schmidt discord $D_{2}$ (FIG.6).
\subsection{Conclusions}
We have studied local time evolution of quantum correlations given
by geometric quantum discord in the system of independent two -
level atoms interacting with environment at zero temperature. The
dynamics is induced by the process of spontaneous emission and we
have local dynamics when only one atom emits and the other is
isolated from the environment. Within this model we have compared
the properties of Hilbert - Schmidt distance discord $D_{2}$ and
trace distance  discord $D_{1}$. When the only local generation of
discord in the classically correlated initial states is considered,
$D_{1}$ and $D_{2}$ provide the similar information: for a large
class of initial states with zero discord and local evolution of
measured subsystem, $D_{1}$ as well as $D_{2}$ grows from zero to
some maximal value and then decay to zero. Moreover, local evolution
of unmeasured subsystem in both cases gives the same result: $D_{1}$
and $D_{2}$ are equal to zero. Local quantum operations can also
increase the existing discord. We have shown that this phenomenon
occurs also in our model of local evolutions  for a large class of
initially discordant states. In contrast to the previous case, now
the behavior of $D_{1}$ and $D_{2}$ significantly differ. First of
all $D_{2}$ increases under the local evolution of unmeasured
subsystem. It happens even in such cases when it decreases under the
evolution of a measured subsystem, which is manifestly nonphysical.
Moreover, $D_{2}$ can increase during the evolution of a measured
subsystem whereas at the same time $D_{1}$ decreases. All those
properties of $D_{2}$ suggests that it is not a reasonable measure
of quantum correlations - more promising is to use trace distance
discord $D_{1}$.


\begin{thebibliography}{99}
\bibitem{Z} H. Ollivier and W.H. Zurek, Phys. Rev. Lett.
\textbf{88}, 017901(2001)
\bibitem{F} A. Ferraro \textit{et al.} Phys. Rev. A \textbf{81},
052318(2010)
\bibitem{Str} A. Streltsov, H. Kampermann and D. Bruss, Phys. Rev.
Lett. \textbf{107}, 170502(2011)
\bibitem{Hu1} X. Hu \textit{et al.} Phys. Rev. A \textbf{84},
022113(2011)
\bibitem{DVB} B. Daki\'{c}, V. Vedral and C. Brukner, Phys. Rev.
Lett. \textbf{105}, 190502(2010)
\bibitem{Piani} M. Piani, Phys. Rev. A \textbf{86}, 034101(2012)
\bibitem{Tuf} T. Tufarelli \textit{et al.} Phys. Rev. A \textbf{86},
052326(2012)
\bibitem{Paula} F.M. Paula, Thiago R. de Oliveira and M.S. Sarandy,
Phys. Rev. A \textbf{87}, 064101(2013)
\bibitem{Cic} F. Ciacarello, T. Tufarelli and V. Giovannetti, New J. Phys.
\textbf{16}, 013038(2014).
\bibitem{Ci} F. Ciccarello and V. Giovannetti, Phys. Rev. A
\textbf{85}, 010102(2012)
\bibitem{Ci1} F. Ciccarello and V. Giovannetti, Phys. Rev. A
\textbf{85}, 022108(2012)
\bibitem{Ge} M. Gessner, E.-M. Laine, H.-P. Breuer and J. Piilo,
Phys. Rev. A \textbf{85}, 052122(2012)
\bibitem{Ca} S. Campbell \textit{et al.} Phys. Rev. A \textbf{84}, 052316(2011)
\bibitem{GJ} M. Gw\'{o}\'{z}d\'{z} and L.Jak\'{o}bczyk, Quantum Inf.
Process. \textbf{13}, 171(2014).
\bibitem{Adesso} D. Girolami and G. Adesso, Phys. Rev. A
\textbf{84}, 052110(2011)
\bibitem{Ozawa} M. Ozawa, Phys. Lett. A \textbf{268}, 158(2000).
\bibitem{Mizra} H. Mehri-Dehnavi, B. Mizra, H. Mohammadzadeh and R. Rahimi, Ann. Phys. \textbf{326}, 1320(2011)
\bibitem{Agarwal} G.S. Agarwal, \textit{Quantum Statistical Theories
of Spontaneous Emission and their Relation to Other
Approaches},Springer, Berlin, 1974.
\end{thebibliography}
\end{document}